\documentclass[10pt,twocolumn,prl,aps,floatfix,longbibliography,superscript
address]{revtex4-1}

\usepackage{color}
\usepackage{siunitx}
\usepackage{graphicx}
\usepackage{amsthm}
\usepackage{amsmath}
\usepackage{amssymb}
\usepackage{enumerate}
\usepackage[table]{xcolor}
\usepackage{booktabs}

\usepackage{pgffor}
\usepackage{pdfpages}
\makeatletter
\AtBeginDocument{\let\LS@rot\@undefined}
\makeatother

\renewcommand{\vec}[1]{\boldsymbol{#1}}

\AtBeginDocument{
\heavyrulewidth=.08em
\lightrulewidth=.05em
\cmidrulewidth=.03em
\belowrulesep=.65ex
\belowbottomsep=0pt
\aboverulesep=.4ex
\abovetopsep=0pt
\cmidrulesep=\doublerulesep
\cmidrulekern=.5em
\defaultaddspace=.5em
}

\begin{document}

\title{Theory of liquid film growth and wetting instabilities on graphene}
\author{Sanghita Sengupta} 
\affiliation{Department of Physics, University of  Vermont, Burlington, VT 05405}
\affiliation{Materials Science Program, University of  Vermont, Burlington, VT 05405}

\author{Nathan S. Nichols} 
\affiliation{Department of Physics, University of  Vermont, Burlington, VT 05405}
\affiliation{Materials Science Program, University of  Vermont, Burlington, VT 05405}

\author{Adrian Del Maestro} 
\affiliation{Department of Physics, University of  Vermont, Burlington, VT 05405}
\affiliation{Materials Science Program, University of  Vermont, Burlington, VT 05405}
\affiliation{Institut f\"ur Theoretische Physik, Universit\"at Leipzig, D-04103, Leipzig, Germany}

\author{Valeri N. Kotov}
\affiliation{Department of Physics, University of  Vermont, Burlington, VT 05405}
\affiliation{Materials Science Program, University of  Vermont, Burlington, VT 05405}

\begin{abstract}
We investigate wetting phenomena near graphene within the
Dzyaloshinskii-Lifshitz-Pitaevskii theory for light gases of hydrogen, helium
and nitrogen in three different geometries where graphene is either affixed to
an insulating substrate, submerged or suspended.  We find that the presence of
graphene has a significant effect in all configurations.  When placed on a
substrate, the polarizability of graphene can increase the strength of the
total van der Waals force by a factor of two near the surface, enhancing the
propensity towards wetting.  In a suspended geometry unique to two-dimensional
materials where graphene is able to wet on only one side, liquid film growth
becomes arrested at a critical thickness which may trigger surface
instabilities and pattern formation analogous to spinodal dewetting.  The
existence of a mesoscopic critical film with a tunable thickness provides a
platform for the study of a continuous wetting transition as well as
engineering custom liquid coatings.  These phenomena are robust to some
mechanical deformations and are  also universally present in doped graphene and
other two-dimensional materials, such as monolayer dichalcogenides.
\end{abstract}

\maketitle


The wetting of an electrically neutral solid surface by a liquid is controlled
by the relative size of attractive van der Waals interactions between molecules
in the liquid and those between the liquid and substrate. For weak
liquid-substrate interactions, the surface may undergo \emph{partial wetting}
manifest as the coexistence of distinct liquid droplets with an atomically thin
layer of adsorbed molecules between them.  In the opposite \emph{complete
wetting} regime, the liquid atoms are strongly attracted to the surface
resulting in the formation of a macroscopically thick film in equilibrium with
the vapor above it \cite{Bonn:2009ha}.  

The growth and stability of this film beyond a few atomic layers is dominated
by the long range \emph{tail} of the van der Waals (vdW) interaction which can
be thought of as creating an effective repulsion between the liquid-vapor and
liquid-substrate boundaries \cite{Dzyaloshinskii:1961vc, Barash:1975hv,
Dzyaloshinskii:2012vn}.  For intermediate liquid-surface interactions it is
possible that at a critical film thickness, $d_c$, (larger than any atomic
length scale) this repulsion vanishes and wetting is arrested due to the lack
of any energetic gain for molecules in the vapor to adsorb into the liquid --
a scenario known as \emph{incomplete wetting} \cite{Combescot:1981wd,
Dash:1982gx, Bonn:2009ha, Huse:1984kf}.

While a phase transition between partial and complete wetting driven by
temperature is generically first order (being controlled by short-distance
details of the adsorption potential) a transition from incomplete to complete
wetting can be continuous due to the presence of only long range vdW forces
\cite{Dietrich:1985ij,Dietrich:1991jx} (critical wetting).  However,
engineering substrates with weak interactions to observe incomplete wetting and
any associated critical phenomena has been challenging, with experiments
concentrating on quantum fluids at low temperatures \cite{Migone,Cheng:1993zza}
or liquid substrates such as alkynes on water at high temperature
\cite{Nakanishi:1982en,Bertrand:2002wa,Rafai:2004hz}.  

\begin{figure*}[t!]
\begin{center}
\includegraphics[width=1.0\textwidth]{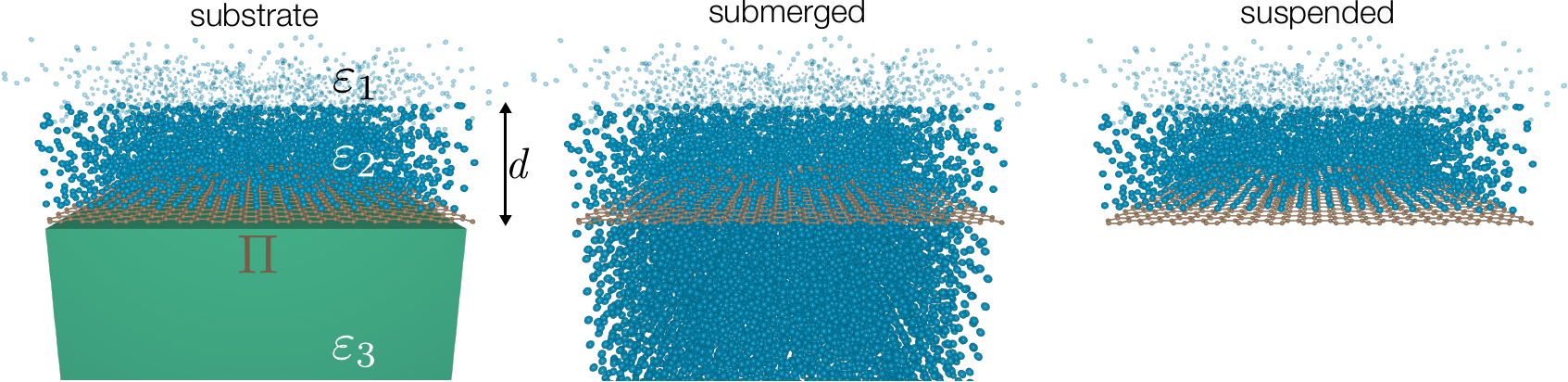} 
\end{center}
\caption{Three geometries that are unique to wetting on two-dimensional
materials.  From left to right -- Substrate: graphene with a momentum and
frequency dependent electronic polarization $\Pi(\vec{q}, i\omega)$ is placed on top of
an insulating substrate with dielectric constant $\varepsilon_3$ and a
macroscopic liquid film with $\varepsilon_2$ grows to a thickness $d$ that is
in equilibrium with its vapor ($\varepsilon_1 \approx 1$).  Submerged: graphene
is floated on top of a liquid with dielectric constant $\varepsilon_2$ and a
liquid film of the same substance grows on the top side. Suspended: a liquid
film grows on top of a pensile graphene sheet.}
\label{fig:wetting}
\end{figure*}

In this letter we report on the physics of wetting in the novel class of geometries
depicted in Fig.~\ref{fig:wetting}, made possible by the ability to readily
fabricate and manipulate atomically flat two-dimensional (2D) crystals such as
graphene \cite{Antonio}, transition-metal dichalcogenides  \cite{Di:2012cs}
(\emph{e.g.}~MoS$_{2}$) and representatives of the 2D topological insulator
family \cite{Ezawa2012, PhysRevB.86.161407, PhysRevB.85.075423, Germanene}
(silicene and germanene). This includes graphene placed on a substrate,
submerged in a liquid, or suspended with a vacuum underneath, realizable due to
the impermeability of graphene to even small atoms \cite{McEuen,Nair442}.

%
%

We devise an extension of the Dzyaloshinskii-Lifshitz-Pitaevskii (DLP) theory
\cite{Dzyaloshinskii:1961vc, Barash:1975hv} (the standard many-body approach
used for accurate analysis of experiments \cite{Sabisky:1973hd, krim}) to
include the polarization of a 2D material in an anisotropic layered dielectric
sandwich.  The results indicate that light gases near suspended 2D materials
are an ideal system to study and characterize critical wetting phenomena.  Our
main findings include: (I) the presence of graphene on a substrate can enhance
liquid film growth consistent with studies of its ``partial wetting
transparency'' to liquid water \cite{Parobek:2015bz}. (II) This effect rapidly
decreases with film thickness and occurs at $\si{\nano\meter}$ scales as
opposed to the $\si{\micro\meter}$ distances where relativistic effects may
become important \cite{Cheng:1988fi}.  (III) In the suspended geometry, the
existence of vacuum beneath graphene causes incomplete wetting with a critical
film thickness on the order of \SIrange{3}{50}{\nano\meter} that can be tuned
through the dynamic polarizability of the adsorbant or the properties of the
semimetal (\emph{e.g.} strain).  This phenomenon  is universally present, and
can be additionally controlled in doped graphene  as well as in insulating
dichalcogenides, thus spanning a wide range of 2D  Dirac materials.  (IV) The
mesoscopic film may exhibit critical surface instabilities including pattern
formation in analogy to spinodal decomposition \cite{Sharma, Reiter, Vrij,
Jain, Herminghaus916, Mitlin, Langer197153}.  Together these findings represent
not only the introduction of a new platform for the study of wetting and
associated critical phenomena, but hint at applications including the creation
of tunable surface coating or drying mechanisms vie electrostatic gating or
mechanical manipulation.

The remainder of this paper is organized as follows: we review 
the DLP theory and show how it is modified by the insertion of a graphene sheet.  We
report quantitative results for wetting and film growth in the three
configurations in Fig.~\ref{fig:wetting}.  For the suspended geometry,
we examine the spreading of droplets on the liquid surface and discuss the
formation of long-wavelength surface instabilities.  We conclude with a
discussion of the experimental measurement of these effects.
Accompanying supplemental materials (SM) provide information on the effects
of temperature, different 2D materials and substrates, uniaxial strain and
electronic doping \cite{supplemental}.

The starting point is the calculation of  the vdW energy $U(d)$ of a charge
neutral system composed of three  substances (having dielectric
functions $\varepsilon_{1,2,3}$) as shown in Fig.~\ref{fig:wetting}, with the
atomically thin graphene layer, characterized by polarization $\Pi$, inserted
at the boundary between regions 2 and 3. $U(d)$ represents the vdW
interaction between the 1-2 and 2-3 material surface boundaries separated by
distance $d$. It is well-known \cite{Israelachvili, Barash:1975hv} that
$U(d)$ can be related to the momentum ($\vec{q}$) and frequency
($\omega$) dependent effective dielectric function ${\cal{E}}({\bf{q}},
i\omega)$ which characterizes the screening of the interlayer Coulomb
potential.  $U(d) = ({\hbar}/{n)(2\pi)^{-3}}\int d^2{\bf q} \int_{0}^{\infty}
d\omega \ln{{\cal{E}}({\bf{q}}, i\omega)}$, where $n=N/V$ is the density of the
liquid (material 2).  It should be noted that for a single-material system
(i.e.  characterized by only one dielectric constant) this formula is simply
the random phase approximation (RPA) correlation energy, while in the case of
anisotropic layered structures it represents the fluctuation (vdW) energy. We
set $\hbar = 1$ from now on.

The calculation of ${\cal{E}}$ involves the electrostatics of a three layer
system. For example, for the configurations of interest in Fig.~\ref{fig:wetting}
one obtains the following formula \cite{Katsnelson, Peeters} for the properly
screened interlayer Coulomb potential $U_{12}$
between 1 and 2: $U_{12} = V_{12}/\varepsilon_g$,  $V_{12}= {8 \pi
e^2 \varepsilon_2}/[{qD(q)}]$, where 
$q = |{\bf q}|$ is the magnitude of the in-plane momentum, 
\begin{equation}
D(q) = (\varepsilon_1 + \varepsilon_2) (\varepsilon_2 + \varepsilon_3) e^{qd} +
(\varepsilon_1 - \varepsilon_2) (\varepsilon_2 - \varepsilon_3) e^{-qd}
\label{eq:Dq}
\end{equation}
and the effect of graphene is in the additional screening characterized by
\begin{equation}
\varepsilon_g({\bf{q}}, i\omega) = 1 - V_2(q)\Pi({\bf{q}}, i\omega). 
\end{equation}
Here, $V_2$ is the Coulomb potential within the lower boundary plane 
\begin{equation}
V_2 = \frac{4 \pi e^2}{qD(q)} \left [ (\varepsilon_1 + \varepsilon_2)e^{qd} +
(\varepsilon_2 - \varepsilon_1)e^{-qd} \right ].
\label{eq:V2}
\end{equation}
The polarization of graphene $\Pi({\bf{q}}, i\omega)$ is 
described in the SM \cite{supplemental, Kotov}.  Then,
keeping in mind that $U_{12} \propto  {e^2 e^{-qd}}/[{q {\cal{E}}({\bf{q}},
    i\omega)}]$, we obtain ${\cal{E}}({\bf{q}}, i\omega) = \varepsilon_g({\bf{q}},
i\omega) D(q) e^{-qd}$, and finally
\begin{equation}
U(d) = \frac{1}{n(2\pi)^{3}} \int d^2{\bf q} \int_{0}^{\infty}  d\omega  \ln{[
\varepsilon_g({\bf{q}}, i\omega) D(q) e^{-qd}]}\, . 
\label{eq:Ud}
\end{equation}

It is instructive to simplify Eq.~(\ref{eq:Ud}) in the limit $(\varepsilon_2 -1)
\ll 1$, which is satisfied with high accuracy for the low-density systems
we have studied (such as He and other light elements). In this case, their vapor can be
considered as vacuum ($\varepsilon_1=1$), and suppressing $\vec{q}$ and
$\omega$ dependence:
\begin{equation}
U(d) \approx \frac{1}{n(2\pi)^{3}}\int d^2{\bf q} \int_{0}^{\infty} d\omega  ({\cal{U}}_d + {\cal{U}}_g),
\label{totalvdw}
\end{equation}
with dielectric:
\begin{equation}
{\cal{U}}_d  = \frac{(\varepsilon_2 -1)(\varepsilon_3 - \varepsilon_2 )}{(\varepsilon_2 +1)(\varepsilon_3 + \varepsilon_2 )}e^{-2qd}
\label{dielectricvdw}
\end{equation}
and graphene parts: 
\begin{equation}
{\cal{U}}_g  =  \frac{ \left (\frac{-4 \pi e^2 \Pi}{q  (\varepsilon_2
    +\varepsilon_3)} \right) \left (\frac{\varepsilon_2 -1}{\varepsilon_2 +1}
\right ) \left (\frac{2\varepsilon_2 }{\varepsilon_2 +\varepsilon_3} \right )
}{ 1 -\frac{4 \pi e^2 \Pi}{q  (\varepsilon_2
+\varepsilon_3)} } e^{-2qd}.
\label{graphenevdw}
\end{equation}
The corresponding vdW force can be obtained from $F(d) = - {\partial  U(d)
}/{\partial d}$ which has dimensions of energy due to the normalization factors
chosen in Eq.~(\ref{eq:Ud}).

When graphene is absent (${\cal{U}}_g=0$), we recover the well
known DLP theory expression \cite{ Dzyaloshinskii:1961vc, LL9,
Dzyaloshinskii:2012vn}.  In particular it describes the important property of
vdW repulsion (a force per unit area known as the disjoining pressure) for
$(\varepsilon_3 - \varepsilon_2)> 0$. We note that inserting graphene will
always lead to repulsion  as $\Pi<0$.
Eq.~(\ref{graphenevdw}) can be used to describe the
three main configurations: graphene on a substrate (characterized by
$\varepsilon_3$),  submerged ($\varepsilon_3= \varepsilon_2$), and
suspended graphene ($\varepsilon_3=1$).

%
\begin{figure}[t]
\begin{center}
\includegraphics[width=1.0\columnwidth]{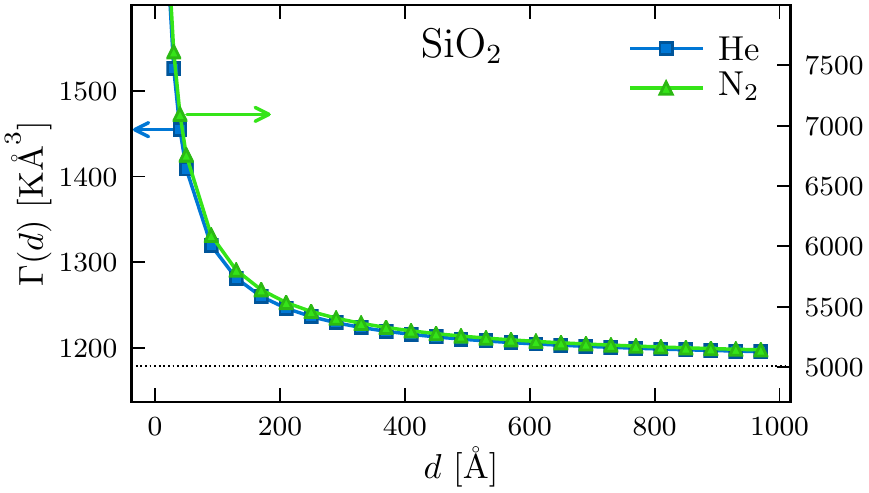}
\end{center}
\caption{Additional liquid film thickness dependence $\Gamma(d)$ (beyond
$1/d^3$) of the van der Waals force between the substrate-liquid and
liquid-vapor interfaces due to the insertion of graphene on a SiO$_2$ substrate. 
The dashed line represents the substrate contribution (in the absence of
graphene) and a crossover from $1/d^4$ to $1/d^3$ is observed. Left vertical
scale corresponds to helium and the right to nitrogen films.}
\label{fig:gasOnSiO2}
\end{figure}
%

To calculate $\mathcal{U}_d$, we take the dielectric function of light
elements to have a single oscillator form: $\varepsilon_{2}(i\omega)  = 1 +
{C_A}/{[1+ (\omega/\omega_A)^2]}$, where for $^4$He we use 
$\omega_A = \omega_{\rm He} \approx \SI{27}{\eV}$ and $C_{\rm He} = 0.054$.
Parameters for other materials are given in the SM \cite{supplemental}.
The substrate dielectric function can typically be well fitted to the form
\cite{Bergstrom}, $\varepsilon_{3}(i\omega)  = 1 + {C_{IR}}/{[1+
(\omega/\omega_{IR})^2]}  + {C_{UV}}/{[1+ (\omega/\omega_{UV})^2]}$.  For
example in the case of SiO$_2$ (quartz):  $\omega_{UV} \approx \SI{13.37}{\eV},
\ \omega_{IR} \approx \SI{0.138}{\eV}$, and $C_{IR} = 1.93, \ C_{UV} = 1.359$.
Other cases are studied in the SM \cite{supplemental}.

The final result can be conveniently written as:
\begin{equation}
F(d) = \frac{\omega_{A}}{n 16 \pi^2}\frac{I(d)}{d^3} \equiv
\frac{\Gamma(d)}{d^3}, 
\label{eq:Fd}
\end{equation}
where the dimensionless expression for $I(d)$ is given in the SM
\cite{supplemental} and is used to calculate $\Gamma(d)$.  It is
clear from Eqs.~(\ref{totalvdw})--(\ref{graphenevdw}) that the
dielectric part leads to a pure $1/d^3$ dependence of the force ($\Gamma(d) =
\Gamma_0$ as $\varepsilon_{1,2,3}$ do not depend on momentum). However, the
graphene contribution has substantial momentum dependence (due to the
polarization $\Pi({\bf{q}}, i\omega)$), and causes a $1/d^4$ law above some
length-scale.  The overall behavior has the scaling form (second term due
to graphene):
\begin{equation}
\Gamma(d)=\Gamma_0 + \frac{\Gamma_1}{d+L}.
\label{eq:Gammad}
\end{equation}
%

%
\begin{figure}[t]
\begin{center}
\includegraphics[width=1.0\columnwidth]{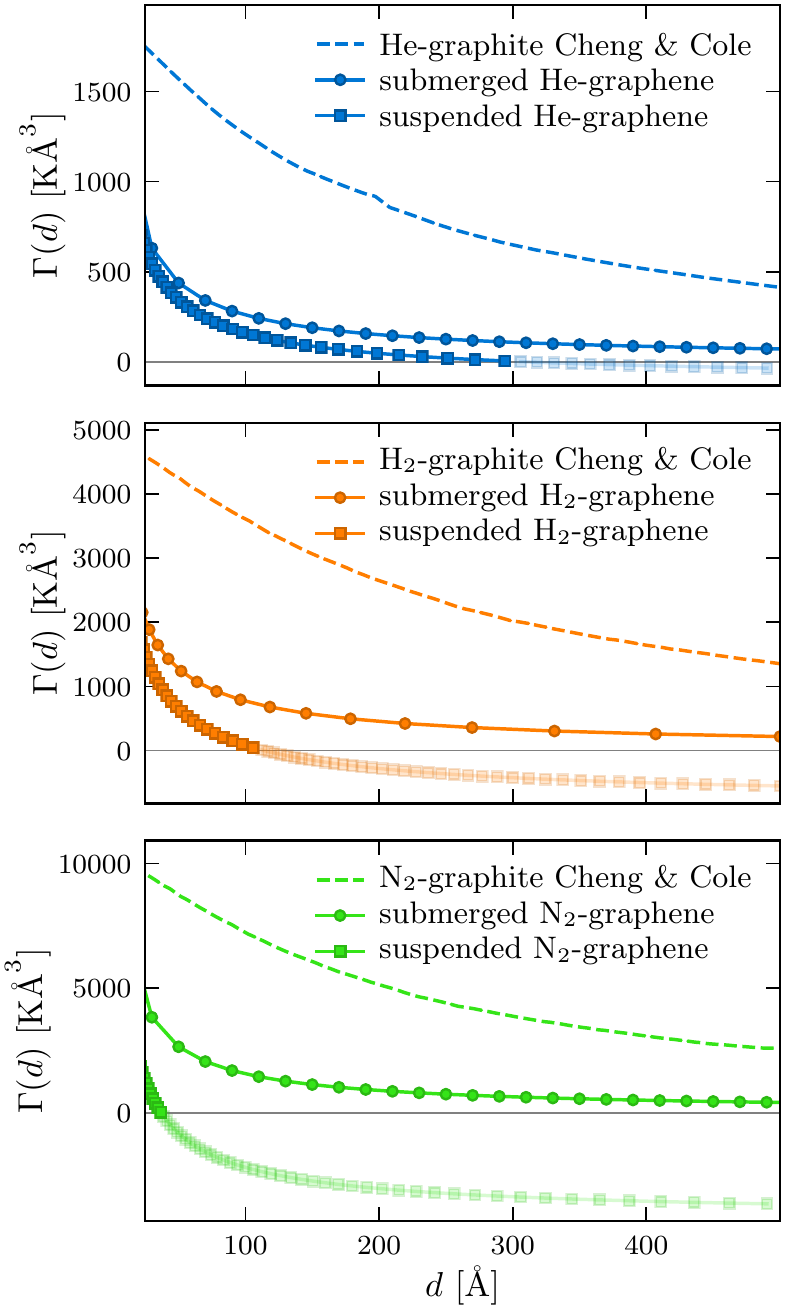}
\end{center}
\caption{Thickness dependence of the van der Waals interaction
    $\Gamma(d)$ for films formed on submerged and suspended graphene (see
    Fig.~\ref{fig:wetting}) composed of helium, hydrogen and nitrogen. The
    dashed line corresponds to films on graphite taken from Cheng and Cole
    \cite{Cheng:1988fi}.  For submerged graphene, $\Gamma(d\to\infty) = 0$
    and in the suspended geometry, there is an instability
    causing film growth to be arrested where $\Gamma(d\ge d_c)
    \le 0$.}
\label{fig:gammad}
\end{figure}
%

Fig.~\ref{fig:gasOnSiO2} shows how the insertion of graphene on a quartz
substrate enhances the vdW repulsion between the substrate-liquid and
liquid-gas interfaces for helium and nitrogen gas.
Graphene introduces a substantial distance dependence to the force that is
larger than that previously reported for graphite \cite{Cheng:1988fi,Li}.
While relativistic corrections can create crossovers in the distance
dependence \cite{Cheng:1988fi}, they happen at larger micron-scales,
while here we see a dominant, purely non-relativistic contribution
at nanometer lengths.  Similar behavior is observed for other substrates such
as 6H-SiC (see SM \cite{supplemental}).  The crossover length $L$
introduced in Eq.~(\ref{eq:Gammad}) is also sensitive to the details of the
substrate and for helium we find that $L \sim \SI{10}{\angstrom}$,
\emph{i.e.}~the crossover toward pure $1/d^4$ behavior in the graphene part occurs
quite rapidly.

%
\begin{figure}[t]
\begin{center}
\includegraphics[width=1.0\columnwidth]{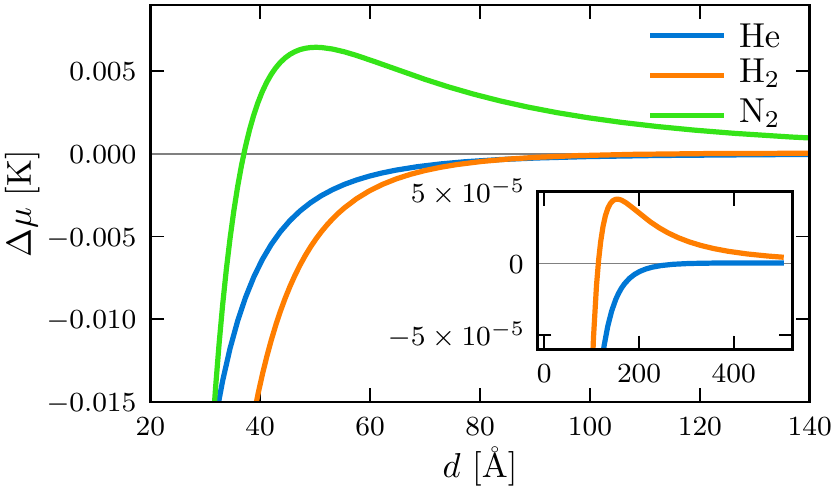}
\end{center}
\caption{Chemical potential of the liquid film (in reference to the bulk)
$\Delta\mu$ as a function of distance $d$ for suspended graphene. An
instability corresponding to $\Delta \mu > 0$ is found for helium, hydrogen
and nitrogen at a finite value of $d=d_{c}$. Inset: zoomed in region of the
main panel showing $d_c$ for He and H$_2$. Values of $d_c$ are reported in
Table~\ref{tab:angle}.}
\label{fig:Δμ}
\end{figure}
%

The vdW force in the submerged and suspended geometries that are unique to 2D
materials can also be evaluated with the results shown in
Fig.~\ref{fig:gammad} for helium, hydrogen and nitrogen films. 
In all cases we compare with calculations from Cheng and Cole \cite{Cheng:1988fi}
for adsorption on graphite (dashed lines).  For submerged graphene ($\varepsilon_3 =
\varepsilon_2$, filled circles) $\mathcal{U}_d = 0$ (see Eq.~(\ref{dielectricvdw})) and
$\Gamma(d)$ decays to zero, in stark contrast to the case of a graphene plated
substrate.  For suspended graphene ($\varepsilon_3 = 1$, squares), we observe a
novel physical effect for all elements:  there is a critical distance
$d_{c}$ at which graphene's (always positive) contribution becomes so weak
it can no longer compensate the negative dielectric part and $\Gamma(d_c) = 0$.
Such an effect is only possible for purely 2D materials that can be suspended
without a supporting substrate - graphene \cite{McEuen, bolotin, meyer} is the
best (but not only) candidate in this family.  For $d > d_c$ the 
liquid film growth stops under equilibrium conditions and the system becomes
unstable. This is the incomplete wetting scenario discussed in the
introduction.  For $d<d_{c}$ the characteristic isotherms that determine the
change of the chemical potential of the film (relative to bulk), $\Delta \mu  =
\mu(d) - \mu(d=\infty)$, are determined by the usual equilibrium condition
(where $P_{0}$ is the saturated vapor pressure) \cite{LL5, Sabisky:1973hd}
\begin{equation}
\Delta \mu = -\frac{\Gamma(d)}{d^3} = T \ln\frac{P}{P_0}, \  \  \   P  \leq
P_{0} .
\label{eq:Δμ}
\end{equation}
Fig.~\ref{fig:Δμ} shows the resulting chemical potential for helium, hydrogen
and nitrogen on suspended graphene which exhibits textbook behavior \cite{LL5} for an
unstable system.  
The suspended film transition from stable $(d < d_c)$ through a metastable
region with $d>d_{c}$ where $\partial \Delta \mu/\partial d >0$; and finally
becomes unstable for $d>d_c, \partial \Delta \mu/\partial d <0$.
The values of the critical film thickness $d_c$ are found to be on the order of
\SIrange{3}{30}{\nano\meter} and are reported in Table~\ref{tab:angle}.

We now concentrate on the properties and implications of the incomplete wetting
scenario where a liquid film with thickness $d_c$ is absorbed on suspended
graphene. As processes governing the further wetting (partial or complete) of
the liquid surface are governed only by the long range tail of the vdW
interaction, they can display a wealth of phenomena of both theoretical and
experimental importance \cite{deGennes, Bonn:2009ha, Huse:1984kf,
    Dietrich:1991jx, Rafai:2004hz,
Ragil:1996eo, BrochardWyart:1991up, Peierls:1978fd, Dash:1982gx, Dash:1977ia,
Nair442, McEuen, Migone, Weiss}.
We can formulate an important question regarding wetting of the liquid film
via a calculation of the contact angle $\theta$ of droplets which can form its
surface.  As these droplets are ``far'' from the substrate, the short-range
adsorption potential is irrelevant, opening up the possibility of universal and
continuous critical behavior. The value of the contact angle is related to the
area under the $\Delta \mu(d> d_c)$ curve \cite{Dzyaloshinskii:1961vc}: 
\begin{equation}
1 - \cos (\theta )  = \frac{n}{\sigma_{l-v}} \int_{d_{c}}^{\infty} \Delta \mu(l)  d l =
 -\frac{n}{\sigma_{l-v}} \int_{d_{c}}^{\infty}  \frac{\Gamma(l)}{l^3}  d l
 \label{eq:ψ}
 \end{equation}
where $\sigma_{l-v}$ is the liquid-vapor surface tension.  Results are shown in
Table~\ref{tab:angle} and we find small angles on the order of a degree that
increase with the polarizability of the adsorbant vapor.
\begin{table}[t]
\begin{center}
    \renewcommand{\arraystretch}{1.5}
    \setlength\tabcolsep{8pt}
  \begin{tabular}{@{}llll@{}} 
   \toprule
   \textbf{Atom} & He & H$_2$ & N$_2$\\ 
    \midrule
    $d_{c}$ (\si{\angstrom})  & $300$ & $120$ & $35$  \\ 
    $\theta\;\ $ (\si{\degree})  & $0.33$ & $0.83$ & $2.41$  \\
    \bottomrule
  \end{tabular}
\end{center}
\caption{\label{tab:angle} Critical film thickness and contact angles for three
    elements.  The surface tensions were taken to be: $\sigma_{He} \simeq
    \SI{0.26}{\milli\newton\per\meter}$, $T=\SI{2.5}{\kelvin}$; $\sigma_{H_2}
    \simeq \SI{2}{\milli\newton\per\meter}$, $T=\SI{20}{\kelvin}$; $\sigma_{N_2}
\simeq \SI{10}{\milli\newton\per\meter}$,  $T=\SI{70}{\kelvin}$.}
\end{table}
The fact that $\theta > 0$ in all cases allows us to consider a remarkable
analogy between surface film instabilities and the
theory of spinodal decomposition \cite{Sharma, Reiter, Vrij, Jain,
Herminghaus916, Mitlin, Langer197153}. The characteristic pattern instability
length scale is governed by the competition between destabilizing vdW forces
and the stabilizing action of the surface tension.  The wavelength $\lambda$ which
corresponds to amplified surface fluctuations (which could ultimately cause
``spinodal dewetting") in the unstable region ($\partial \Delta \mu/\partial d
<0$) is given by (for $d \gg L$)
\begin{equation}
\lambda^2 \simeq  -8\pi^2\frac{\sigma_{l-v}}{n\left(\frac{\partial \Delta \mu}{\partial d}\right)}
 \approx \frac{8\pi^2}{3}\frac{\sigma_{l-v} d^4}{n|\Gamma_0|}.
 \label{eq:λ}
\end{equation}
From Fig.~\ref{fig:gammad}, for example for H$_2$ we can estimate
$|\Gamma_0| \sim 10^3\ \si{\kelvin\angstrom^3}$, which yields 
$\lambda \sim 10^4-10^5\ \si{\angstrom}$ for $d \approx 150-300\ \si{\angstrom}$.

In conclusion we have considered how the relatively weak van der Waals
interactions between light atoms and graphene can substantially affect their
wetting behavior when graphene is placed on a substrate, submerged in a liquid
or suspended above vacuum. We find that placing graphene on a substrate
enhances its propensity towards wetting during initial film growth
which may have implications for its use as a conductive coating.  For suspended
graphene, the absence of any substrate material leads to an instability where
film growth becomes arrested at a critical thickness.  As the vapor pressure
above this film is increased, droplets may form, driving surface fluctuations
which can potentially have large amplitudes.  It is significant that  the
critical film thickness $d_c$ is  dependent on mechanical deformations (e.g.
uniaxial strain) in graphene, and is also universally present for other  2D
materials, such as members of   the group-VI dichalcogenides family  (MoS$_2$,
WS$_2$, MoSe$_2$, etc.) \cite{supplemental}.  
Quite importantly, we also find that the instability occurs in doped graphene,
within a wide range of experimentally accessible carrier densities
\cite{supplemental}. Thus we conclude that this is a universal
phenomenon in suspended 2D Dirac materials, ranging from insulating monolayer
dichalcogenides to semi-metallic (undoped)  and doped graphene.  The exact
value of $d_c$ itself,  which we find to be on the order of several hundred Angstroms,
depends on material characteristics such as band gap, quasiparticle velocity,
strain and doping level. Experimental confirmation of
these effects would involve the measurement of adsorbed film thickness using
standard quartz microbalance \cite{Migone, VanCleve2008} or interferometry
\cite{Rafai:2004hz} techniques.  The ability to electronically or mechanically
manipulate free-standing atomically flat substrates opens up the possibility of
producing an exotic quantum wetting phase transition driven by a non-thermal
control parameter.

We acknowledge stimulating conversations with O. Sushkov, M. Cole, and P.
Taborek on the general subject of van der Waals forces and wetting.  N. Nichols
was supported in part by the Vermont Space Grant Consortium under NASA Grant
and Cooperative Agreement NNX15AP86H.  A.D.~acknowledges the German Science Foundation (DFG) for financial support via grant RO 2247/10-1.

\bibliography{refs}



\section*{Supplementary material}

# Supplementary material for "Theory of liquid film growth and wetting instabilities on graphene"

**Sanghita Sengupta, Nathan Nichols, Adrian Del Maestro, Valeri Kotov**

In this supplement we provide additional information and details on the theory of Dzyaloshinskii-Lifshitz-Pitaevskii, the polarization function of graphene and the dielectric constants of the light gases and substrates studied in the main text. We provide complementary results for different substrates in general experimental use and present additional calculations on how the conclusions of the main text would be altered when placing the two-dimensional (2D) substrate under uniaxial strain or doping. Results for several transition-metal monolayer dichalcogenides (which exhibit energy gaps) are compared with graphene. We conclude with a technical discussion on the effects of temperature which can be safely neglected in the parameter regime of interest in the main text. In what follows, we adopt the units $\hbar = k_B = 1$.

### Dzyaloshinskii-Lifshitz-Pitaevskii (DLP) theory

The DLP theory [S1, S2] is the standard many-body approach used to calculate the van der Waals forces in a substrate-liquid-vapor configuration and then relate the film thickness to pressure when analyzing experiments (Eq.(10)). Comparison with experiments on various materials [S3–S5] generally shows very good agreement, with deviations occurring in various circumstances typically attributable to presence of forces beyond long-range van der Waals (e.g. short-range forces), or the very complex nature of dielectric screening, etc. From a theoretical viewpoint DLP provides calculation of the van der Waals effective force between the two boundaries ("disjoining pressure"). It treats the dielectrics as effective homogeneous media with certain frequency-dependent dielectric functions, which themselves could be hard to calculate but are fairy well-known for standard materials after many years of comparison with experiment. The polarization loops within DLP theory are basically re-summed self-consistently in close analogy with the RPA (random phase approximation), which is the standard way to include self-consistent screening. It should also be noted that examination of different theoretical aspects of DLP theory [S1, S6, S7] shows that the approach used in the main text of the paper (based on the calculation of the effective dielectric function) is equivalent to other approaches, such as those based on determining electromagnetic fluctuations via photon Green's functions in a medium and the van der Waals stress tensor.

### Graphene Polarization and Gas Parameters

The dynamical polarization of graphene at zero chemical potential (charge neutrality point), on the imaginary frequency axis, has the form [S8]:

$$\Pi(\mathbf{q}, i\omega) = -\frac{1}{4} \frac{|\mathbf{q}|^2}{\sqrt{v^2|\mathbf{q}|^2 + \omega^2}}, \tag{S1}$$

where $v = 6.6$ eV Å is the velocity of the Dirac quasiparticles.

The parameters of the three light gases (medium 2) discussed in the main text are as follows (see e.g. [S9, S10]). For Helium the dynamical dielectric constant is

$$\varepsilon_2(i\omega) = 1 + 4\pi n_{\mathrm{He}} \alpha(i\omega), \quad \alpha(i\omega) = \frac{\alpha_{\mathrm{He}}}{1 + (\omega/\omega_{\mathrm{He}})^2}, \tag{S2}$$

where the density $n_{He} = 2.12 \times 10^{-2}$ Å$^{-3}$, the static polarizability $\alpha_{He} = 1.38$ a.u., and the characteristic oscillator frequency $\omega_{He} = 27.2$ eV. The atomic unit of polarizability is defined as 1 a.u. $= 0.148$ Å$^3$.

For Nitrogen and Hydrogen, which have densities comparable to Helium but significantly larger polarizabilities, more accurate formulas based on the Clausius-Mossotti relation are typically used:

$$\varepsilon_2(i\omega) = 1 + \frac{4\pi n_A \alpha(i\omega)}{1 - \frac{4\pi}{3} n_A \alpha(i\omega)}, \quad A = \mathrm{N}_2, \mathrm{H}_2, \tag{S3}$$

The dynamical polarizability is defined as in Eq. (S2). For H$_2$ the parameters are: $n_{H_2} = 2.04 \times 10^{-2}$ Å$^{-3}$, $\alpha_{H_2} = 5.44$ a.u., $\omega_{H_2} = 14.09$ eV. For N$_2$: $n_{N_2} = 1.73 \times 10^{-2}$ Å$^{-3}$, $\alpha_{N_2} = 11.74$ a.u., $\omega_{N_2} = 19.32$ eV.

A convenient expression for the dimensionless quantity $I(d)$ appearing in Eq. (8) is

$$I(d) = \int_0^\infty \int_0^\infty dy dx \, x^2 \, e^{-x} \frac{\varepsilon_2(y) - 1}{\varepsilon_2(y) + 1} \frac{1}{\varepsilon_2(y) + \varepsilon_3(y)} \times$$
$$\left\{ \varepsilon_3(y) - \varepsilon_2(y) + \frac{4gx\varepsilon_2(y)}{[\varepsilon_2(y) + \varepsilon_3(y)] \sqrt{x^2 + y^2 [\Omega(d)]^2} + 2gx} \right\}, \tag{S4}$$

where $\Omega(d) = 2\omega_A d/v$ and $x = 2qd$. Since $\varepsilon_2(i\omega)$ is a function of the frequency ratio, $\varepsilon_2(i\omega/\omega_A)$, we define $y = \omega/\omega_A$ and use the notation $\varepsilon_2(i\omega/\omega_A) \to \varepsilon_2(y)$. In the substrate dielectric function $\varepsilon_3$, see below, this leads to the rescaling $\varepsilon_3(y) = 1 + \frac{C_{IR}}{1 + y^2(\omega_A/\omega_{IR})^2} + \frac{C_{UV}}{1 + y^2(\omega_A/\omega_{UV})^2}$. The dimensionless coupling $g = (\pi/2)(e^2/v) \approx (\pi/2)(2.2)$ characterizes the Coulomb interaction strength in graphene in vacuum.

### Substrate Effects

It is instructive to investigate the effect of different substrates (medium 3) on wetting. Many dielectric substrates are accurately described by the formula [S11]

$$\varepsilon_3(i\omega) = 1 + \frac{C_{IR}}{1 + (\omega/\omega_{IR})^2} + \frac{C_{UV}}{1 + (\omega/\omega_{UV})^2}. \tag{S5}$$

As an example, we have performed calculations for 6H-SiC, see Fig. S1, to be compared with our results in Fig. 2 of the main text (for a SiO$_2$ substrate). While the magnitude of the van der Waals force shows significant variations between substrates, the overall distance dependence, and crossover length scale remains comparable.

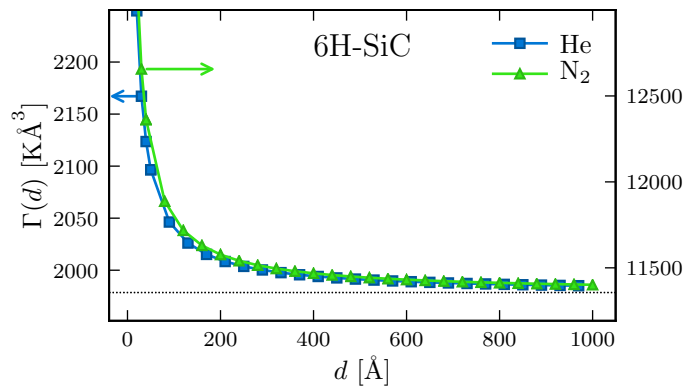

FIG. S1. Van der Waals force contribution $\Gamma(d)$ for a 6H-SiC substrate with parameters $C_{IR} = 3.67$, $\omega_{IR} = 0.1 \, \text{eV}$, $C_{UV} = 5.53$, $\omega_{UV} = 7.39 \, \text{eV}$ [S11]. This can be compared with that of a SiO$_2$ substrate plotted as Fig. 2 of the main text.

Another important quantity which characterizes the van der Waals force is the length scale $L$ defined in Eq. (9) which sets the crossover from $1/d^3$ to $1/d^4$ behavior. Due to the Dirac fermion motion in graphene (i.e. the strong momentum dependence of graphene's polarization) the force crosses over to a stronger power law at distances beyond $L$. In Fig. S2 we present a general analysis of the substrate dependence of $L$, for He. Due to helium's large atomic frequency $\omega_{He} \approx 27 \, \text{eV}$, it is easy to see that only the ultraviolet (UV) part in Eq. (S5) provides an important contribution to the relevant integral, while the infrared (IR) part is irrelevant. Our results can be used to characterize a variety of substrates [S11], beyond the two main ones studied in this work. It is important to notice that $L$ remains fairly small (several Å) for practically all available substrates.

### Influence of Uniaxial Strain in Graphene

The existence of two-dimensional (2D) materials such as graphene opens up the attractive possibility to manipulate van der Waals forces by mechanical manipulation of the substrate – strain. This is due to the strong influence of

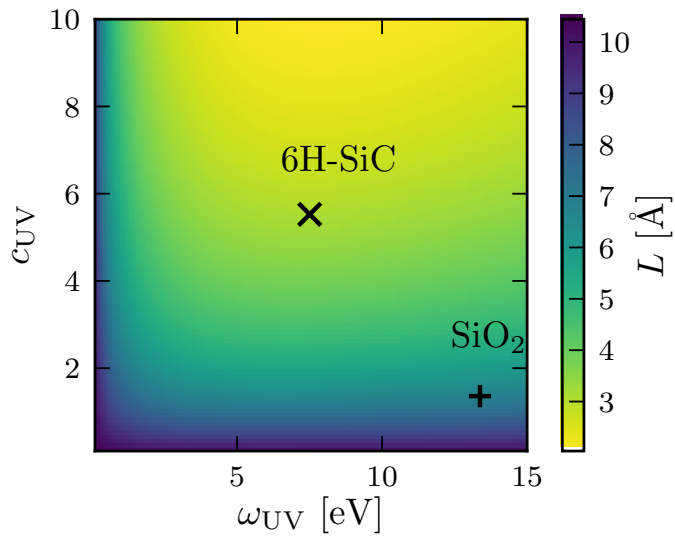

FIG. S2. Map of the crossover length $L$ for He, defined in Eq. (9), as a function of the two main substrate parameters.

strain on the electronic motion and consequently substantial change in graphene's polarization [S12, S13]. The case of uniaxial strain is the most straightforward to analyze, in which case graphene's polarization is

$$\Pi(\mathbf{q}, i\omega) = -\frac{1}{4v_x v_y} \frac{v_x^2 q_x^2 + v_y^2 q_y^2}{\sqrt{v_x^2 q_x^2 + v_y^2 q_y^2 + \omega^2}}. \tag{S6}$$

Here we assume strain is in the $y$ (armchair) direction leading to decrease of the electron velocity $v_y$ in that direction (while the velocity in the perpendicular ($x$) direction remains practically unchanged). It is convenient to introduce the ratio $v_\perp = v_y/v_x < 1$ which reflects the strain (relative increase in lattice spacing); this ratio is perturbatively proportional to strain for small values but exhibits non-linear behavior for larger deformations [S12, S13]. For example $v_\perp = 0.2$ corresponds to 34% strain, $v_\perp = 0.4$ corresponds to 25% strain, and $v_\perp = 0.75$ corresponds to 10% strain.

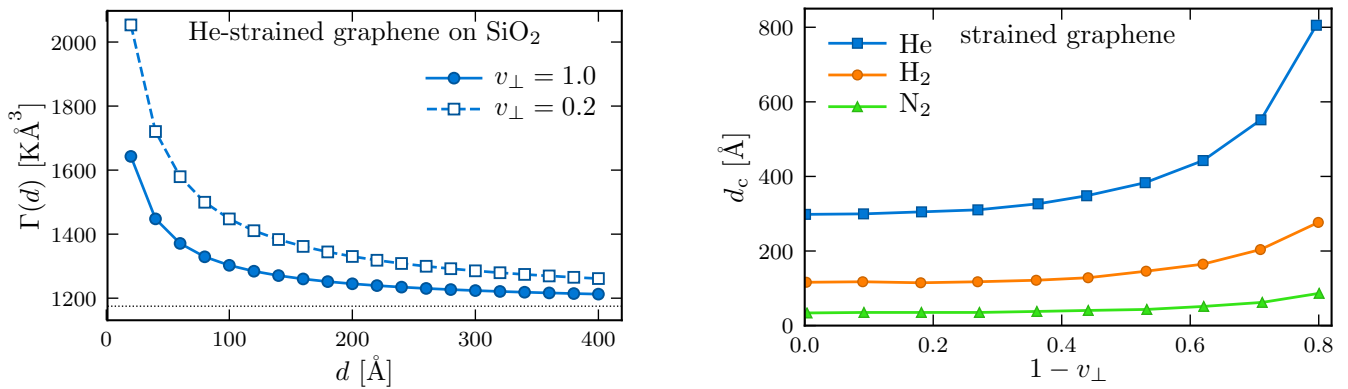

FIG. S3. Left: Influence of strong (for illustrative purposes) graphene strain ($v_y/v_x = 0.2$, 34% strain) for He on SiO$_2$ substrate, leading to significant enhancement of the van der Waals force. The value $v_y/v_x = 1$ corresponds to isotropic, unstrained graphene. Right: Strain dependence of critical film thickness $d_c$ in the suspended graphene geometry. For example the parameter value $1 - v_\perp = 0.8$ corresponds to 34% strain, while $1 - v_\perp = 0.6$ corresponds to 25% strain. The general tendency, particularly important for moderate to strong strain, is promotion of film growth (increase of critical $d_c$).

Figure S3 quantifies the effects of strain. Due to the increase of the graphene polarization Eq. (S6), the van der Waals interaction increases and thus strain promotes wetting. While at present such strong strain is difficult to achieve

in graphene, mechanical deformations are also expected to be present in a variety of 2D materials [S14], and thus the general tendencies described here could be important in a variety of physical situations.

As a consequence of the enhanced van der Waals interaction with strain, the tendency towards film growth is enhanced. In particular, while in the suspended geometry (suspended graphene with no supporting substrate) films can grow only up to a finite thickness $d_c$, the value of $d_c$ increases with strain as illustrated in Fig. S3 (right). This effect is quite weak for small strain and becomes significant as strain grows. In addition we find substantial dependence on the type of atom (the effect is strongest for helium).

### 2D Materials: Insulating Dichalcogenides and Doped Graphene

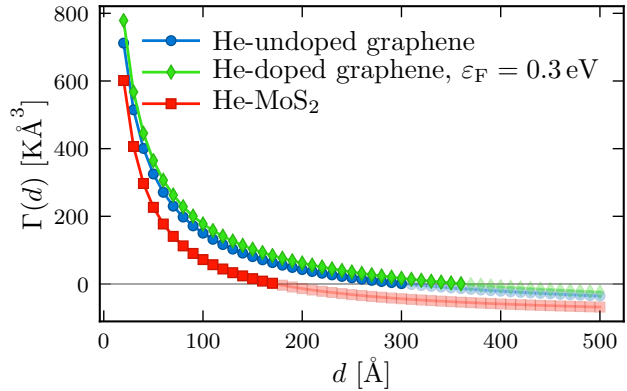

FIG. S4. Variation of the van der Waals force $\Gamma(d)$ in the case of He in the suspended geometry: $MoS_2$ compared to doped and undoped graphene. The result shows significant decrease of the maximum film thickness $d_c$ (from 300Å to $\approx$ 180Å), caused by the suppression of the force due to the presence of the large electronic gap $\Delta$ in $MoS_2$. A representative (quite substantial) doping value is also shown to illustrate the tendency with doping. The shown value of $\varepsilon_F$ corresponds to graphene carrier density $n_e \approx 7 \times 10^{12}$ cm$^{-2}$, well above the lowest-possible density that can be achieved in suspended graphene ($\approx 10^{9-10}$ cm$^{-2}$). More detailed variation with density appears on the next figure.

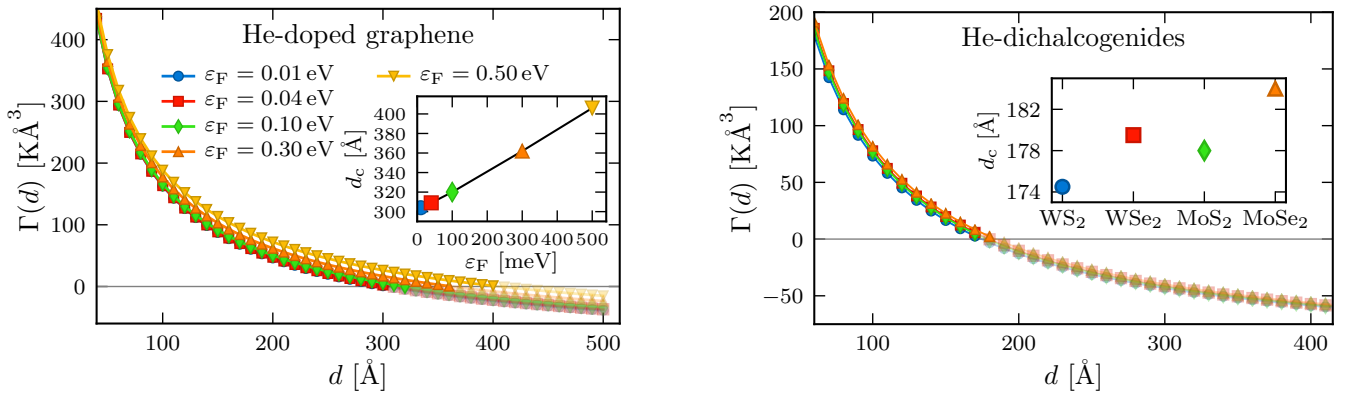

FIG. S5. Left: Variation of the van der Waals force $\Gamma(d)$ in the case of He in the suspended geometry, for doped graphene. The densities corresponding to the shown Fermi energies are: $n_e \approx 8 \times 10^9, 10^{11}, 10^{12}, 7 \times 10^{12}, 2 \times 10^{13}$ cm$^{-2}$. For smaller densities ($\varepsilon_F < 40$ meV) the results are indistinguishable from the undoped situation. The Fermi energy is related to the density via $\varepsilon_F = v\sqrt{\pi n_e}$. The overall effect of doping is quite small and becomes noticeable for higher densities only. The general tendency is an increase of $d_c$ due to the increased polarization of doped graphene (inset). Right: Comparison of various members of the dichalcogenides family, showing remarkably similar behavior. The material parameters, $\Delta$ – gap, $v$ – velocity, $g = (\pi/2)e^2/v$ – Coulomb interaction, are: $WS_2$: $\Delta = 1.79$ eV, $v = 4.38$ eV Å, $g = 5.16$; $WSe_2$: $\Delta = 1.6$ eV, $v = 3.94$ eV Å, $g = 5.74$; $MoS_2$: $\Delta = 1.66$ eV, $v = 3.51$ eV Å, $g = 6.44$; $MoSe_2$: $\Delta = 1.47$ eV, $v = 3.11$ eV Å, $g = 7.26$.

*Dichalcogenides.* While graphene provides the most well studied example of a 2D material, it is important to assess the applicability of our results to other 2D compounds. Numerous 2D materials have been discovered, forming groups suitable for designing so-called van der Waals heterostructures [S15]. In particular the group-VI dichalcogenides [S16] include e.g. $MoS_2$, $WS_2$, $MoSe_2$, etc., which exhibit a significant electronic gap $\Delta$ of order 1 eV, as well as a small spin-orbital interaction. For the purpose of van der Waals calculations, the most significant modification (compared to graphene) to be taken into account is the presence of the gap, while the spin-orbital component can be neglected. The polarization function in this case is [S17]

$$\Pi(\mathbf{q}, i\omega) = -\frac{|\mathbf{q}|^2}{\pi}\left[\frac{m}{\tilde{q}^2} + \frac{1}{2\tilde{q}}\left(1 - \frac{4m^2}{\tilde{q}^2}\right)\tan^{-1}\left(\frac{\tilde{q}}{2m}\right)\right], \quad \tilde{q} \equiv \sqrt{v^2|\mathbf{q}|^2 + \omega^2}, \quad m = \Delta/2, \tag{S7}$$

where $m$ is the Dirac mass (half of electronic gap). For example for $MoS_2$ $\Delta = 1.66$ eV, and other materials have similar parameters [S16]. Neglecting the small spin-orbit interaction can result in several percent error but our main conclusions will remain intact. We assume materials are in their insulating phases, i.e. the Fermi energy is in the gap.

We consider the wetting instability in the suspended geometry, as was done for graphene in the main text (Figure 3). The main effect of the gap is the suppression of the van der Waals force. Consequently this causes a significant enhancement of the instability (film growth arrested at smaller thickness $d = d_c$) as shown in Fig. S4 for $MoS_2$, to be compared with results for graphene in Figure 3 of the main text. Results for other dichalcogenides are similar and summarized in Fig. S5 (right). Therefore the critical wetting instability is present in the four main members of the dichalcogenides family, with very similar values of the critical film thickness $d_c$.

*Doped Graphene.* It is also important to investigate the effect of finite carrier concentration $n_e$ in graphene, when the Fermi energy $\varepsilon_F$ is shifted away from the Dirac (charge neutrality) point, and compare with results for undoped graphene ($\varepsilon_F = 0$). Experimentally, in suspended graphene samples [S18, S19], the carrier density can be very small $n_e < 10^{10}$ cm$^{-2}$, and thus a close proximity to the Dirac point can be achieved, with $\varepsilon_F \sim 10$ meV.

We use the well-known polarization for doped graphene [S20]

$$\Pi(\mathbf{q}, i\omega) = -\frac{q^2}{4\sqrt{\omega^2 + v^2 q^2}} - \frac{2\varepsilon_F}{\pi v^2} + \frac{q^2}{2\pi\sqrt{\omega^2 + v^2 q^2}}\Re e\left[\arcsin\left(\frac{2\varepsilon_F + i\omega}{vq}\right) + \left(\frac{2\varepsilon_F + i\omega}{vq}\right)\sqrt{1 - \left(\frac{2\varepsilon_F + i\omega}{vq}\right)^2}\right] \tag{S8}$$

and our results are summarized in Fig. S5 (left). It is clear that at small densities the results for doped and undoped graphene are practically identical. Significant modification of $d_c$ starts appearing only in the regime $\varepsilon_F > 100$ meV, which corresponds to substantial density $n_e \gtrsim 10^{12}$ cm$^{-2}$. It is natural that the instability tends to become suppressed with doping (i.e. $d_c$ tends to increase) since the polarization of graphene in the metallic regime increases. However we can confidently conclude that in the low-density regime, easily achievable in suspended graphene samples, our results discussed in the main body of the paper remain practically unchanged. In addition, the critical wetting instability also remains fully present in strongly-doped suspended graphene (Fig. S5 (left).)

Overall we conclude that the instability in the suspended geometry, with $d_c$ within several hundred Angstroms, is present in all three main 2D Dirac material groups: (insulating) dichalcogenides, semi-metallic (undoped) graphene, and doped graphene. Figure S4 summarizes the main tendencies exhibited by representatives of those groups, namely an increase of the critical film thickness as systems transition from insulating to metallic behavior.

### Effect of Temperature

The instability in the suspended graphene geometry is essentially a zero temperature phenomenon, since we work in the temperature range where the light elements we consider form a liquid (e.g. $T \approx 2$K for He, and somewhat higher $T \approx 20$K for H$_2$). In fact we can readily see that the effect of temperature at the distances involved in any of our geometries (up to several hundred Angstroms) is negligible. Indeed, the basic expressions Eqs. (6–7) of the main text have the following form, which we now write in the finite temperature formalism, introducing the Matsubara frequencies $\omega_m$ (the prime means that the zero frequency term should be divided by two)

$$T\sum_m{}'\left(f_1(i\omega_m) + f_2(i\omega_m)\frac{e^2\Pi(q, i\omega_m)}{q}\right), \quad \omega_m = 2\pi m T. \tag{S9}$$

Here we have written explicitly only the frequency sums, not showing the momentum integrations (weighted by the exponential factors $e^{-2qd}$). The function $f_1(i\omega_m) \sim [\varepsilon_2(i\omega_m) - 1]^2$ in the suspended geometry ($\varepsilon_3 = 1$), while

$f_1(i\omega_m) \sim [\varepsilon_2(i\omega_m) - 1]$ in the other geometries with substrates present. This is the purely dielectric part. In the second term, involving graphene, we have $f_2(i\omega_m) \sim [\varepsilon_2(i\omega_m) - 1]$, proportional to the atomic polarizability. There exist other combinations of substrate dielectric screening factors within these two functions but we do not write them explicitly since it is easy to see that their existence is harmless and does not change the arguments that follow.

In the first term, since the dielectric function enters via the frequency ratio $\omega_m/\omega_A$ (Eq. (S2)), i.e. the combination $T/\omega_A \ll 1$, it is evident that the sum transforms into the zero temperature frequency integration. Recall that $\omega_A \sim 10\,\mathrm{eV} \sim 10^5\,\mathrm{K}$.

In the second term, the characteristic momentum is of order $q = q^* \sim 1/d$ (which is the momentum where the integral over $q$ accumulates due to the presence of the exponential factor). Taking into account that $\Pi$ has the form Eq. (S1) it is clear that as long as $T \ll v/d$, the sum is essentially at zero temperature. One should keep in mind that: (I) This condition is satisfied up to distances of several hundred Angstroms, e.g. $v/d \approx 150$K at $d = 500$Å, and the ratio increases at smaller distances, (II) the polarization itself has a finite $T$ contribution, i.e. is of the form $\Pi(\mathbf{q}, i\omega_m) + \Pi_T(\mathbf{q}, i\omega_m)$, where $\Pi_T(\mathbf{q}, i\omega_m)$ can be found explicitly [S21]. Calculations that take into account the full temperature dependence of the polarization confirm the above estimates and show that indeed the corrections are small and certainly negligible up to several hundred Å.

To summarize, our main conclusions, especially concerning the instability leading to finite film thickness $d_c \lesssim 400$Å as shown in Figures 3-4 of the main text, remain valid at finite (but small) temperatures which is the regime of interest for liquid phases of light elements. At larger distances, finite temperature as well as relativistic corrections will gradually become more pronounced.

---


\end{document}